\documentclass[12pt]{iopart}

\usepackage{iopams}
\usepackage{setstack}
\usepackage{graphicx}

\begin{document}
\title[Exciton effective mass enhancement in electric and magnetic fields]{Exciton effective mass enhancement in coupled quantum wells in electric and magnetic fields}
\author{J Wilkes and E A Muljarov}
\address{School of Physics and Astronomy, Cardiff University, Cardiff CF24~3AA, UK}
\ead{wilkesj1@cf.ac.uk}
\begin{abstract}
We present a calculation of exciton states in semiconductor coupled quantum wells (CQWs) in the presence of electric and magnetic fields applied perpendicular to the QW plane. The exciton Schr\"odinger equation is solved in real space in three dimensions to obtain the Landau levels of both direct and indirect excitons. Calculation of the exciton energy levels and oscillator strengths enables mapping of the electric and magnetic field dependence of the exciton absorption spectrum. For the ground state of the system, we evaluate the Bohr radius, optical lifetime, binding energy and dipole moment. The exciton mass renormalization due to the magnetic field is calculated using a perturbative approach. We predict a non-monotonous dependence of the exciton ground state effective mass on magnetic field. Such a trend is explained in a classical picture, in terms of the ground state tending from an indirect to a direct exciton with increasing magnetic field.
\end{abstract}
\submitto{\NJP}
\maketitle

\section{Introduction}
\label{intro}
Spatially-indirect excitons in coupled quantum wells (CQWs) present a model system for the study of a statistically degenerate Bose gas in solid-state materials. This is due to the long optical lifetime of indirect excitons in comparison to their thermalization time which permits the creation of a cold and dense exciton gas \cite{ButovJPCM2007,KuznetsovaPRB2012,AlloingPRB2012}. The exciton lifetime depends on the overlap of electron and hole wave functions. For indirect excitons, this can be varied over several orders of magnitude \cite{ButovPRB1999a,SivalertpornPRB2012} by changing the electric field applied perpendicular to the growth direction.

Numerous studies on indirect excitons have focused on Bose-Einstein condensation \cite{HighNANO2012,AlloingEPL2014,HighNAT2012,EisensteinNAT2004}, spatial pattern formations \cite{ButovNAT2002,RemeikaPRB2013,AlloingARXIV2012} and spin transport phenomena \cite{HighPRL2013}. Transport of indirect excitons has been studied extensively. Their dipolar nature provides a probe of the density via a blue shift in the emission \cite{LaikhtmanPRB2009,IvanovPRL2010,CohenPRL2011}. It also facilitates transport of excitons over tens of micrometers \cite{HammackPRB2009} as indirect excitons can screen the QW disorder potential due to interface roughness and defects \cite{IvanovEPL2002}. Moreover, the dipole-orientation of the excitons provides the possibility to control transport via patterned electrodes positioned adjacent to the CQWs \cite{GartnerAPL2006,HighSCI2008,LeonardAPL2012}. Exciton transport has been studied in electrostatic traps \cite{NegoitaPRB1999,RapaportPRB2005,RapaportAPL2006,ChenPRB2006,GartnerPHE2008}, linear potential gradients \cite{GartnerAPL2006,LeonardAPL2012} and stationary \cite{RemeikaPRL2009} and moving \cite{WinbowPRL2011} lattices. The high degree of control led to demonstrations of devices such as excitonic optical transistors \cite{AndreakouAPL2014}. The properties of indirect excitons are also important for understanding a new breed of quasi-particles known as dipolaritons \cite{ChristmannPRB2010}. Such states are realized by embedding CQWs inside a planar Bragg-mirror microcavity \cite{ChristmannAPL2011}. At the resonant tunneling condition, the dipolariton is a superposition of a cavity photon, a spatially-direct and an indirect exciton and thus acquires a static dipole moment \cite{CristofoliniSCI2012,SivalertpornARXIV2013}. Dipolaritons have been suggested for observation of super radiant THz emission \cite{KyriienkoPRL2013} and continuous THz lasing \cite{KristinssonPRB2013}.

The application of magnetic fields to the CQW structure provides an extra degree of control. In exciton systems, magnetic fields are used for trapping \cite{SeidlPRB2011,FreireBJP2002,Rahimi-ImanJPCS2013}, control of superfluidity \cite{LozovikJETP1975,RuboPLA2006}, tuning of exciton resonances \cite{PietkaPRB2015} and studying spin-splitting phenomena \cite{JadczakPRB2012,Rahimi-ImanPRB2011}. The focus of this paper is the exciton effective mass enhancement due to a magnetic field \cite{ButovPRL2001}. From a qualitative view point, this effect can be understood by considering a simple classical analogy. The Lorentz force experienced by two oppositely charged particles that travel perpendicular to a uniform magnetic field acts to increase their relative separation. This comes at an energy cost as it is opposed by the Coulomb attraction between the charges. The energy cost manifests itself as a resistance to change in momentum and thus an effective mass increase, dependent on the applied magnetic field strength. We consider the case where both electric and magnetic fields are normal to the QW plane. By finding a numerically exact solution of the exciton Schr\"odinger equation, we perform a precise calculation of the mass enhancement. This description is an important ingredient for simulating exciton transport kinetics in a magnetic field where the exciton effective mass is a critical parameter \cite{KuznetsovaPRB2012,RemeikaPRB2013,HammackPRB2009,LeonardAPL2012,WinbowPRL2011,WilkesPRL2012}.

Several works (\cite{MoralesPRB2008,LopezSSS2010} for example) used variational methods to study indirect excitons in magnetic fields. However, these all lacked a calculation of the effective mass enhancement. A theoretical study of excitons in CQWs with finite barrier and well width was done in \cite{DzyubenkoPRB1996}, using two-dimensional (2D) free electron-hole pair states in magnetic field as a basis for expansion of the exciton wave function. This approach is only well suited for treating high magnetic fields. Also, a calculation of the exciton mass renormalization was absent. The mass renormalization in the 2D limit was calculated for excitons in single QWs \cite{ArseevJETP1998} and indirect excitons in double QWs \cite{LozovikPRB2002}. In those works, the electron and hole wave functions in the QW growth direction were approximated by delta functions located at the QW positions. This corresponds to the limit of infinitely deep QWs of zero width and is applicable only for the analysis of either purely direct or purely indirect excitons in the case of high electric field and strong QW confinement. It cannot describe in any detail the smooth crossover between direct and indirect exciton states at intermediate electric fields.

The recently developed rigorous multi-sub-level approach \cite{SivalertpornPRB2012} (MSLA) was previously used to calculate the CQW exciton states in perpendicular electric fields. The outcomes, which included a calculation of the dependence of the absorption spectrum and optical lifetime on electric field, were in good agreement with available experimental data. Here, we use the MSLA to solve the exciton Schr\"odinger equation in CQWs with both electric and magnetic fields applied perpendicular to the QW plane. This method solves the exciton in-plane problem by a numerical discretization in real space and is thus equally well suited for both low and high magnetic fields. We have refined the MSLA to a highly accurate scheme by using Numerov's algorithm \cite{Numerov1924}, a fourth-order linear multistep method, for obtaining the energies and wave functions of the exciton states. It is found that the direct-to-indirect exciton crossover with varying electric and magnetic fields is an important ingredient in describing the magnetic field induced exciton effective mass renormalization, which was not captured in earlier works. We note that the MSLA can also be used to solve the exciton Schr\"odinger equation coupled to Maxwell's equations in order to model microcavity polaritons \cite{SivalertpornARXIV2013}. However, modelling polaritons in magnetic fields is beyond the scope of this paper and forms the subject of future study.

In Section \ref{model}, we describe the Schr\"odinger equation for a CQW exciton in the presence of electric and magnetic fields pointing in the growth direction and provide details of the MSLA. The perturbative approach used to obtain the electric and magnetic field dependence of the exciton effective mass is outlined in Section \ref{massRenorm}. In Section \ref{results}, we present calculations of exciton states and their associated properties for 8-4-8 nm GaAs/AlGaAs CQWs. Comparisons with 2D methods are made in Section \ref{2Dcomparison} and we show that our method converges to these in the limit of zero QW thickness. Conclusions are made in Section \ref{conclusions}.

\section{Excitons in multiple QWs in electric and magnetic fields}
\label{model}
To calculate the bound exciton states in a semiconductor nanostructure with external bias and magnetic field orientated normal to the QW plane, we consider the following Hamiltonian:
\begin{equation}
\hat{H}({\bf r}_e,{\bf r}_h) = \hat{H}_e({\bf r}_e) + \hat{H}_h({\bf r}_h) + V_C(|{\bf r}_e - {\bf r}_h|) + E_g,
\label{Ham}
\end{equation}
with
\begin{eqnarray}
\hat{H}_{e,h}({\bf r}) = \hat{p}_{e,h}({\bf r}) \frac{1}{2}\hat{m}^{-1}_{e,h}(z) \hat{p}_{e,h}({\bf r}) + U_{e,h}(z), \label{Heh} \\
\hat{p}_{e,h}({\bf r}) = \frac{\hbar}{i}\nabla_{\bf r} \pm \frac{e}{c}{\bf A}({\bf r}), \\
V_C(r) = -\frac{e^2}{\varepsilon r}.
\end{eqnarray}
Here, ${\bf r}_{e,h}$ are the electron and hole coordinates. For a multi-layered heterostructure, we adopt a cylindrical coordinate system $(\rho,\phi,z)$ with the $z$-axis  along the QW growth direction. The electron and hole effective mass tensors, $\hat{m}_{e,h}(z)$, are composed of in-plane and perpendicular components, $m^{||}_{e,h}$ and $m^{\perp}_{e,h}(z)$, respectively. As in \cite{SivalertpornPRB2012}, we neglect here any $z$-dependence in the in-plane electron and hole effective masses, which is justified by relatively low mass contrast in the heterostructures treated here and a minor contribution to the exciton problem of the electron and hole wave functions outside the well regions. $\hat{m}_{e,h}(z)$ are layer-dependent step functions, as are the QW confinement potentials $V^{QW}_{e,h}(z)$. The confinement potential and the external bias are included in the potentials $U_{e,h}(z) = V^{QW}_{e,h}(z) \pm eFz$ where $F$ is the electric field. For the band gap $E_g$ and in-plane masses $m^{||}_{e,h}$, values for the QW layers are used. $\varepsilon$ is the average permittivity of the structure. For a perpendicular magnetic field (along the growth direction), one takes the vector potential, using the symmetric gauge, in the form ${\bf A}({\bf r}) = \frac{1}{2}{\bf B}\times{\bf r}$.

\begin{figure}
\centering
\includegraphics[width=0.7\textwidth]{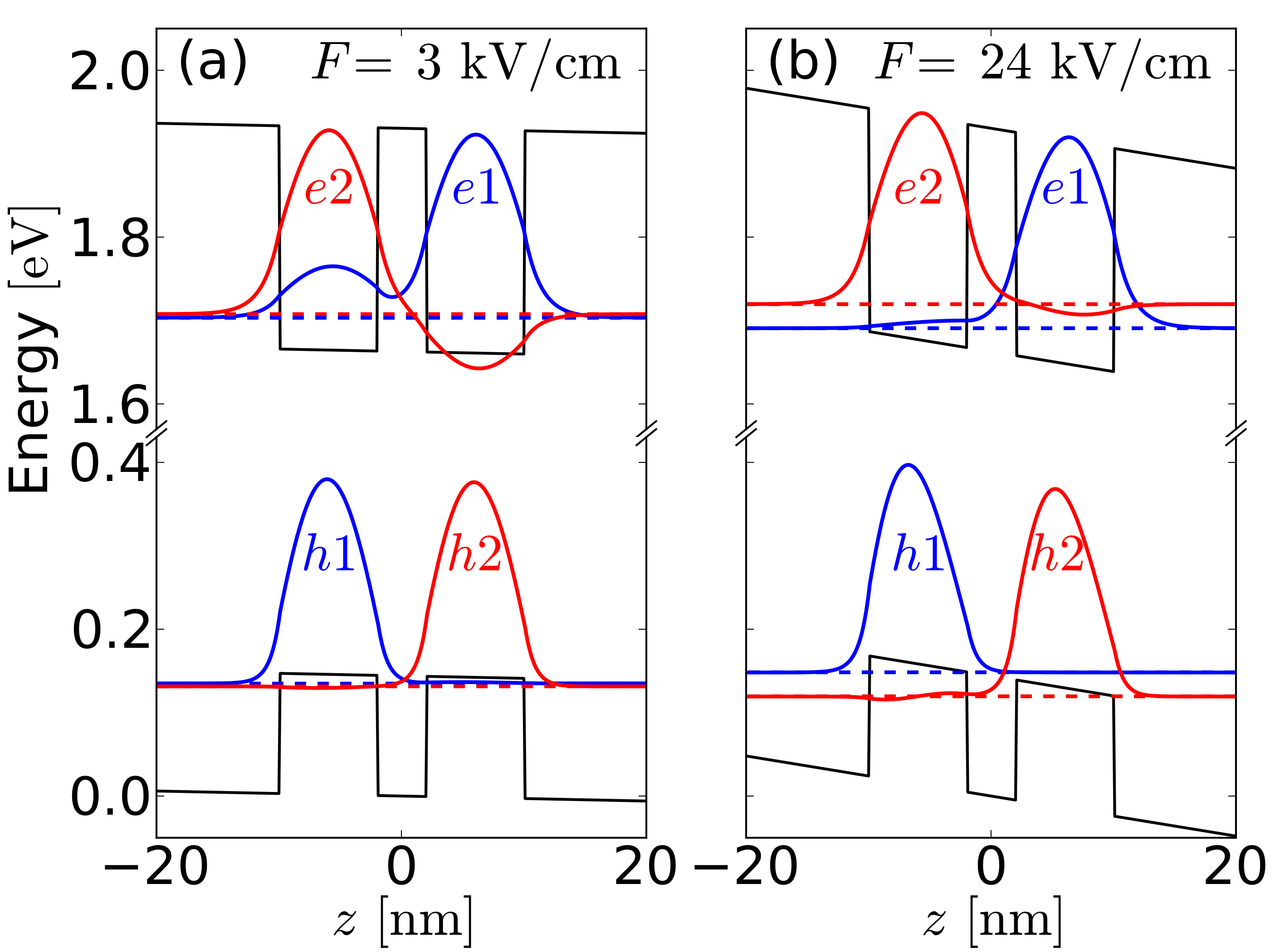}
\caption{Ground ($e1$ and $h1$) and first excited ($e2$ and $h2$) electron and hole states for (a) $F = 3\,{\rm kV/cm}$ and (b) $F = 24\,{\rm kV/cm}$ in 8--4--8-nm $\rm GaAs/Al_{0.33}Ga_{0.67}As$ CQWs.}
\label{zWfn}
\end{figure}

It can be shown \cite{GorkovJETP1968,LozovikJETP1997} that the solution to the equation $\hat{H}_0\Psi = E\Psi$ has the following variable-separable form:
\begin{equation}
\Psi({\bf r}_e,{\bf r}_h) = \chi_{\bf P}(\boldsymbol\rho,{\bf R})e^{im\phi}\varphi(\rho,z_e,z_h)
\label{COM1}
\end{equation}
with the center of mass motion of an exciton carrying 2D momentum ${\bf P}$ described solely by
\begin{equation}
\chi_{\bf P}(\boldsymbol\rho,{\bf R}) = \exp \left\{i\left[{\bf P} + \frac{e}{c}{\bf A}(\boldsymbol\rho) \right] \cdot \frac{{\bf R}}{\hbar}\right\},
\label{COM2}
\end{equation}
where ${\bf R} =\boldsymbol\rho_e m_e^{||}/M_{\rm x} +\boldsymbol\rho_h m_h^{||}/M_{\rm x}$ and $\boldsymbol\rho = \boldsymbol\rho_e - \boldsymbol\rho_h=(\rho,\phi)$ are the 2D in-plane center of mass and relative coordinates, respectively, $M_{\rm x} = m_e^{||} + m_h^{||}$ is the in-plane exciton mass, and $m$ is the exciton magnetic quantum number. Such a substitution enables removal of the center of mass coordinate from the problem. The exciton relative motion is then described by the Hamiltonian $\hat{H}^{\bf P}_{\rm x}$ derived in \ref{appA}. For zero in-plane momentum, the Hamiltonian $\hat{H}^{{\bf P}=0}_{\rm x}$ takes the form
\begin{eqnarray}
\hat{H}^{0}_{\rm x}(z_e,z_h,\rho) = \hat{H}^\perp_e(z_e) + \hat{H}^\perp_h(z_h) + \hat{K}(\rho) + V_B(\rho) \nonumber \\ + V_C\left(\sqrt{\rho^2 + (z_e - z_h)^2}\right) + E_g,
\label{mainHamiltonian}
\end{eqnarray}
where the Hamiltonians of the electron and hole perpendicular motion are
\begin{equation}
\hat{H}^\perp_{e,h}(z) = - \frac{\hbar^2}{2}\frac{\partial}{\partial z}\frac{1}{m^{\perp}_{e,h}(z)}\frac{\partial}{\partial z} + U_{e,h}(z).
\label{Weh}
\end{equation}
These describe the Coulomb-uncorrelated single particle electron and hole states in the absence of a magnetic field, with the wave functions ${\psi}^{e,h}_q(z)$, where the index $q$ labels the electron (hole) states quantized in the CQW heterostructure potentials. The Schr\"odinger equations $\hat{H}^\perp_{e,h}{\psi}^{e,h}_q = {E}^{e,h}_q {\psi}^{e,h}_q$ are solved numerically using the shooting method with Numerov's algorithm. Figure \ref{zWfn} shows the ground and first excited electron and hole states for finite electric fields in a GaAs/AlGaAs CQW.

The kinetic term in (\ref{mainHamiltonian}), $\hat{K}(\rho)$, is given by
\begin{equation}
\hat{K}(\rho) = -\frac{\hbar^2}{2 \mu}\left[\frac{\partial^2}{\partial \rho^2} + \frac{1}{\rho}\frac{\partial}{\partial \rho} - \frac{m^2}{\rho^2} \right],
\label{kinetic}
\end{equation}
where $\mu$ is the in-plane reduced exciton mass, $1/\mu = 1/m_e^\parallel + 1/m_h^\parallel$. The potential due to the magnetic field is
\begin{equation}
V_B(\rho) = \frac{e \hbar m B}{2 \varkappa c} + \frac{e^2 B^2 \rho^2}{8 \mu c^2}.
\label{Bpotential}
\end{equation}
where $1/\varkappa = 1/m_e^\parallel - 1/m_h^\parallel$ is the inverse mass difference of the electron and hole, called the exciton magnetic dipole mass \cite{LozovikPRB2002}.

The Schr\"odinger equation $\hat{H}^{0}_{\rm x} \left| k, m \right\rangle = E_{k,m} \left| k, m \right\rangle$, in which $ \left| k, m \right\rangle$ is an exciton state in a CQW in finite electric and magnetic fields, is solved using the MSLA developed in \cite{SivalertpornPRB2012}. The principle of the method is that we expand the exciton wave function $\varphi(\rho,z_e,z_h)$ in (\ref{COM1}) into the set of Coulomb uncorrelated electron-hole pair states leading to
\begin{equation}
\left| k, m \right\rangle = e^{i m \phi} \sum_{n=1}^{N_eN_h} \Phi_n(z_e,z_h) {\phi}^{k,m}_n(\rho),
\label{solution}
\end{equation}
where  $\Phi_n(z_e,z_h) = {\psi}_{p_n}^e(z_e) {\psi}_{q_n}^h(z_h)$. The mapping $n \mapsto (p_n,q_n)$ leads to $N_eN_h$ pair states formed from $N_e$ electron and $N_h$ hole states. We then calculate the radial components of the wave function ${\phi}^{k,m}_n(\rho)$, using a matrix generalization of the shooting method with Numerov's method incorporated in the finite difference scheme. This solved the problem of the electron and hole in-plane and perpendicular motion mixed by the Coulomb interaction, leading to the exciton quantization labeled here by the index $k$. The numerical solution is generated on a logarithmic grid in the range $\rho \in [\rho_0,R]$. A sufficiently small inner limit $\rho_0$ is chosen so that the wave functions and extracted data have converged with respect to decreasing $\rho_0$. The outer limit $R$ of the wave functions is sufficiently large that all radial components asymptotically approach zero. The existence of such a limit is guaranteed for non-zero $B$ as the potential $V_B$ grows with $\rho^2$. This is in contrast to the case of $B=0$ where truncation of the solution range leads to a discretization of the unbound {\it e-h} continuum states. For the range of electric fields considered ($F < 25\,{\rm kV/cm}$), it was previously found for $B=0$ that high accuracy is achieved using two electron and hole states ($N_e = N_h = 2$) \cite{SivalertpornPRB2012}.

The state with zero angular momentum has radiative linewidth $\Gamma_R^{(k)} = (\pi e^2 \hbar f^{(k)})/(\sqrt{\varepsilon} m_0 c)$. The oscillator strength, $f^{(k)}$, is calculated from the overlap of the electron and hole wave functions,
\begin{equation}
f^{(k)} = \frac{2 m_0 E_{k,0} |d_{cv}|^2}{\hbar^2} \left| \sum_{n=1}^{N_eN_h} {\phi}^{k,0}_n(0) \int \Phi_n(z,z) \, dz \right|^2,
\label{oscStrength}
\end{equation}
where $d_{cv}$ is the dipole matrix element between the conduction and valence bands.

\section{Exciton mass renormalization in a magnetic field}
\label{massRenorm}

For an exciton with a finite center of mass momentum ${\bf P}\neq0$, the Hamiltonian (\ref{mainHamiltonian}) is modified to 
\begin{equation}
\hat{H}_{\rm x}^{\bf P} = \hat{H}_{\rm x}^{0} + \hat{V}_{\bf P}\,, 
\end{equation}
where the perturbation $\hat{V}_{\bf P}$ is given by (see \ref{appA})
\begin{equation}
\hat{V}_{\bf P} = \frac{P^2}{2M_{\rm x}} + \frac{2e}{M_{\rm x}c}{\bf P} \cdot {\bf A}(\boldsymbol\rho)\,,
\label{pert}
\end{equation}
and $P = |{\bf P}|$. To calculate the exciton mass renormalization in a magnetic field, we treat $P$ as a small parameter and use perturbation theory up to 2nd order. The 1st term of (\ref{pert}) contributes to the exciton energy in 1st order only leading to the bare mass. The 2nd term, vanishing in 1st order, gives rise to a quadratic in $P$ correction in 2nd order and is thus responsible for the mass renormalization. Neglecting non-parabolicity of the exciton band, which would be accounted for by treating the 2nd term in higher perturbation orders, we find the correction to the exciton energy proportional to $P^2$ and renormalized effective mass $M^\ast_{k,m}$ of exciton state $\left| k, m \right\rangle$ in magnetic field:
\begin{equation}
\frac{1}{M^\ast_{k,m}} \!=\! \frac{1}{M_{\rm x}} + 2 \!\left (\!\frac{2e}{M_{\rm x}c P}\!\right)^2\! \!\sum_{\underset{s = \pm 1}{j \neq k}} \!\frac{\left| \left\langle k,m\left| {\bf P} \!\cdot\! {\bf A}(\boldsymbol\rho) \right| j,m+s \right\rangle \right|^2}{E_{k,m} - E_{j,m+s}}\,.
\label{effectiveMass}
\end{equation}
Here the index $j$ counts over the eigenstates with the angular momenta $m \pm 1$ which only contribute in 2nd order and which are calculated at given values of electric and magnetic field. Details of the calculation of the matrix elements in (\ref{effectiveMass}) are given in \ref{appB}.

The advantage of our approach compared to some previous calculations of the exciton mass renormalization is that the perturbation is in $P$ only, allowing the $B$-field to be arbitrarily large. In contrast, the approach developed in \cite{ArseevJETP1998} used the magnetic field as a small parameter of the perturbation theory. Consequently, the applicability of the latter is restricted to low magnetic field (up to $\approx 2\,{\rm T}$ in the structure considered in this paper, see also the comparison in Section \ref{2Dcomparison}). Another significant benefit of our approach is that the full 3D solution of the exciton Schr\"odinger equation (\ref{mainHamiltonian}) describes the inter-well coupling that is neglected by any 2D approximation \cite{ArseevJETP1998,LozovikPRB2002}.

\section{Application to GaAs/AlGaAs CQWs}
\label{results}
All results in this section refer to the structures studied in \cite{ButovJPCM2007,KuznetsovaPRB2012,ButovPRB1999a,HammackPRB2009,LeonardAPL2012,RemeikaPRL2009,WinbowPRL2011} which consist of symmetric $\rm GaAs/Al_{0.33}Ga_{0.67}As$ QWs with barrier and well thickness of 4\,nm and 8\,nm, respectively. A complete list of parameters used in the calculations are given in table \ref{parameters}. Although we restrict the present analysis to two QWs, the method is applicable to any number of QWs since one can always obtain, either analytically or numerically, eigenstates of the single particle electron and hole Hamiltonians in (\ref{Weh}).
\begin{table}
\begin{center}
\caption{Parameters of the model}
\begin{tabular}{l l l}
\hline
$\varepsilon$    & Relative permittivity          & 12.5          \\
$E_g$            & GaAs band gap                  & 1.519\,eV     \\
$m^{\perp}_e(z)$ & Electron mass in QW            & 0.0665\,$m_0$ \\
                 & Electron mass in barrier       & 0.0941\,$m_0$ \\
$m^{\perp}_h(z)$ & Hole mass in QW                & 0.34\,$m_0$   \\
                 & Hole mass in barrier           & 0.48\,$m_0$   \\
$M_{\rm x}$      & In-plane exciton mass          & 0.22\,$m_0$   \\
$\mu$            & In-plane reduced exciton mass  & 0.042\,$m_0$  \\
$\varkappa$      & Exciton magnetic dipole mass   & 0.15\,$m_0$   \\
$d_{cv}$         & Dipole matrix element          & 0.6\,nm       \\
\hline
$\rho_0$         & Inner wave function boundary   & 0.025\,nm     \\
$R$              & Outer wave function boundary   & 500\,nm       \\
                 & Number of in-plane grid points & 300           \\
                 & $z$-grid spacing               & 0.1\,nm       \\
\hline
\label{parameters}
\end{tabular}
\end{center}
\end{table}

\begin{figure}[htp]
\centering
\includegraphics[width=0.7\textwidth]{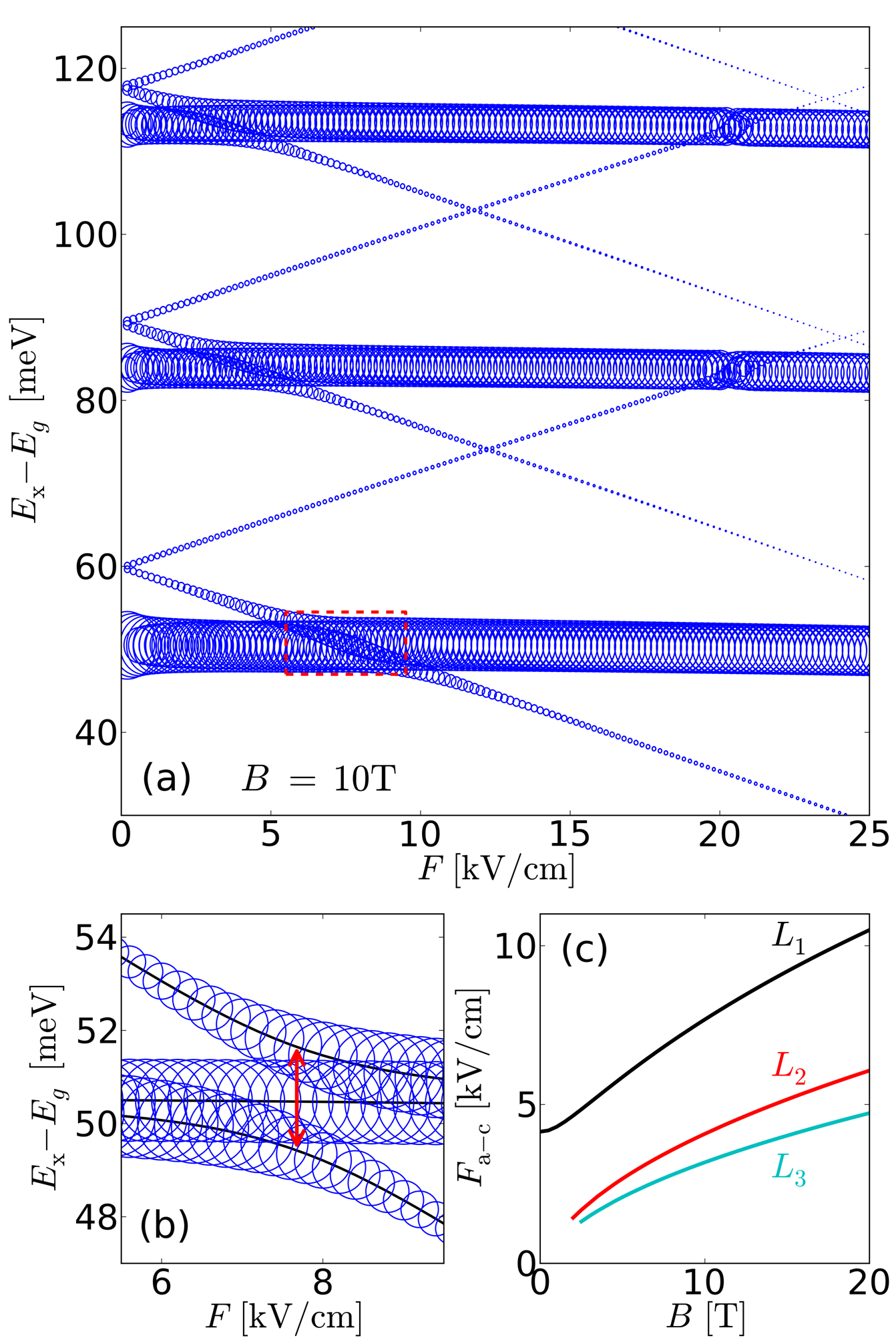}
\caption{(a) Electric field dependence of the optical transition energy for different exciton states. The circle area is proportional to the exciton oscillator strength. (b) Magnified image [indicated in (a) by the boxed area] of the anti-crossing between the direct and indirect exciton states in the first Landau level, $L_1$. Black curves show the exact transition energies. The red arrow shows the minimum energy separation between the lower and upper curves. (c) The electric field, $F_{\rm a\mbox{-}c}$, at which the minimum separation occurs between the upper and lower curves in (b) as a function of $B$.}
\label{LandauLevels}
\end{figure}

Figure \ref{LandauLevels}(a) shows the optical transition energy for different exciton states as a function of electric field at $B=10\,{\rm T}$. For each state, the circle area is proportional to the oscillator strength. One can identify four different exciton states with each state being duplicated in each of the first three Landau levels. Firstly, we see a pair of bright states at $E_{\rm x} - E_g \approx 51\,{\rm meV}$ whose energy is almost independent of electric field. These are the first Landau level ($L_1$) direct exciton states occupying adjacent QWs. The direct exciton states are close in energy and appear almost as a single state in figure \ref{LandauLevels}(a). Secondly, we see a pair of darker states, one whose energy decreases and one whose energy increases with electric field from $E_{\rm x} - E_g \approx 60\,{\rm meV}$ at $F=0$. These are the $L_1$ indirect exciton states. Their transition energy dependence on electric field is almost linear with the gradient corresponding to the nominal distance between the QWs. The state with energy that decreases linearly with $F$ is composed predominantly of the ground electron and hole states e1h1 shown in figure \ref{zWfn}. The other is composed predominantly of the excited electron and hole states e2h2. The corresponding $L_2$ and $L_3$ replicas of these four states appear at higher energies. The spacing between the Landau levels for each type of exciton is approximately equal.

In the previous work using the MSLA \cite{SivalertpornPRB2012}, which did not include magnetic field, it was necessary to calculate a large number of unbound {\it e-h} pair states in order to determine the exciton spectrum. These states were discretized due to a finite domain of the solution which imposed an infinite potential barrier at $\rho = R$. It was found that with increasing $R$, the density of such discretized continuum states increased while their oscillator strengths decreased. The absorption spectrum converged for sufficiently large $R$. In the present case, however, the magnetic field removes any need to consider the continuum states. The potential due to the magnetic field, (\ref{Bpotential}), contains a term in $\rho^2$ which naturally closes the domain of the solutions. Therefore, a finite value of $R$ can always be found such that the wave functions are sufficiently small there. This point explains the apparent cleanliness of figure \ref{LandauLevels}(a) which does not contain any unbound {\it e-h} pair states.

Figure \ref{LandauLevels}(b) shows a magnification in the region of anti-crossing between the $L_1$ direct and indirect exciton states, labeled in figure \ref{LandauLevels}(a). We identify the electric field at the anti-crossing $F_{\rm a\mbox{-}c}$ as the value of $F$ where the minimum separation occurs between the upper and lower black curves. This is shown in figure \ref{LandauLevels}(c) as a function of $B$ for the first three Landau levels. The magnetic field provides a means to tune $F_{\rm a\mbox{-}c}$ over several kV/cm. The corresponding energy splitting between the upper and lower black curves in figure \ref{LandauLevels}(b) is about 2\,meV and varies by less than 10\% in the investigated magnetic field range.

\begin{figure}
\centering
\includegraphics[width=0.7\textwidth]{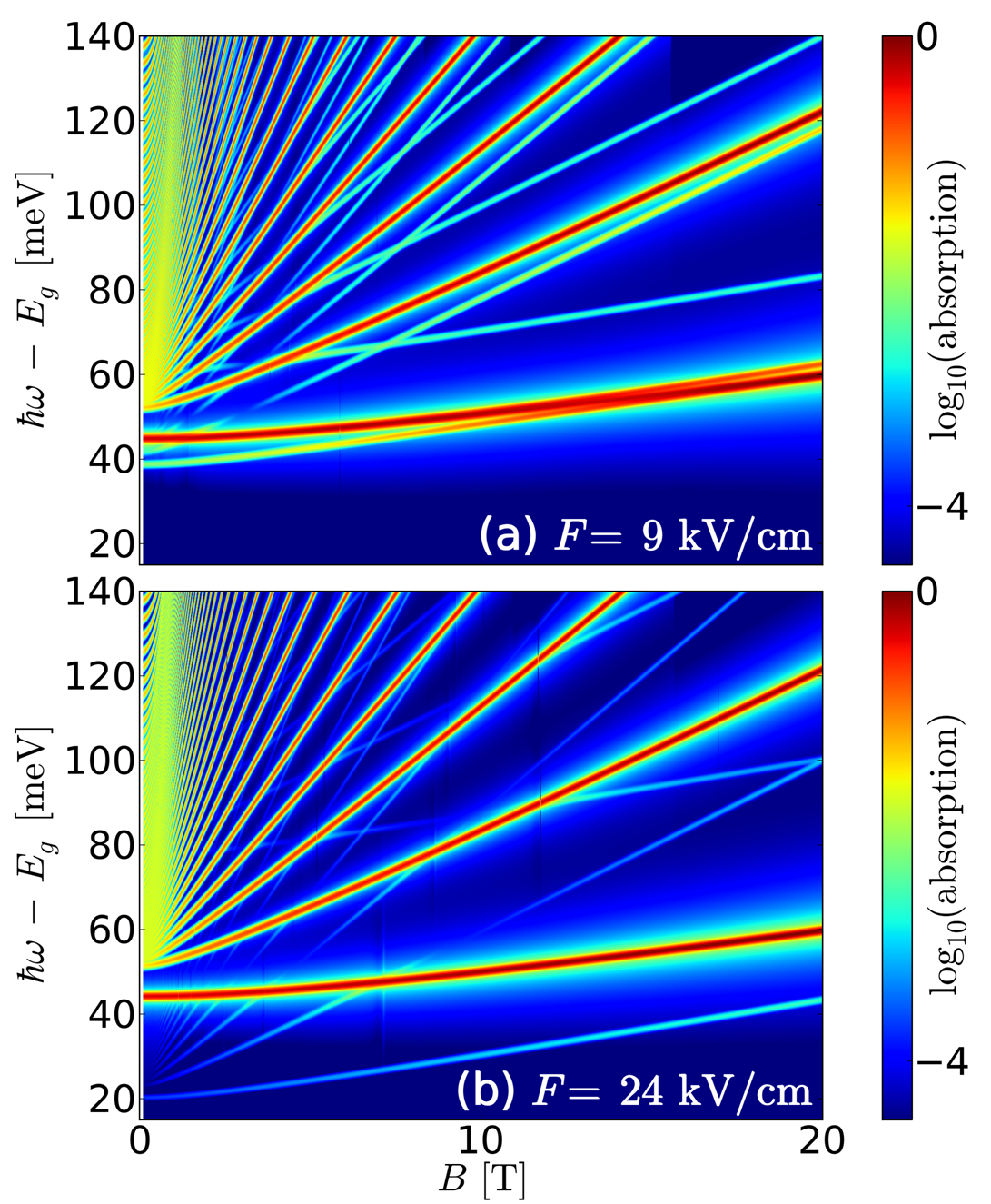}
\caption{Magnetic field dependence of the exciton absorption spectrum at (a) 9\,kV/cm and (b) 24\,kV/cm. The spectra are convoluted with a Gaussian of width 1meV. Both spectra are normalized to 1 at the maximum.}
\label{spectrum}
\end{figure}

Using the Lorentzian model of absorbing oscillators, the absorption spectrum is calculated from the transition energy and linewidth of each exciton state. The magnetic field dependence of the absorption spectrum is plotted in figure \ref{spectrum}. For clarity, the spectrum has been broadened by convolution with a Gaussian of full width at half maximum of 1\,meV. Most clearly visible is the Landau {\it fan} of the direct exciton states. These bright spectral lines are due to the two direct exciton states overlapped in energy. Fainter fans are seen for the indirect exciton states -- one starting from about 20\,meV and another from about 80\,meV in figure \ref{spectrum}(b). The lower (upper) fan corresponds to indirect excitons composed mainly of an electron and a hole in the ground (excited) state, e1h1 (e2h2).

\begin{figure}
\centering
\includegraphics[width=1.0\textwidth]{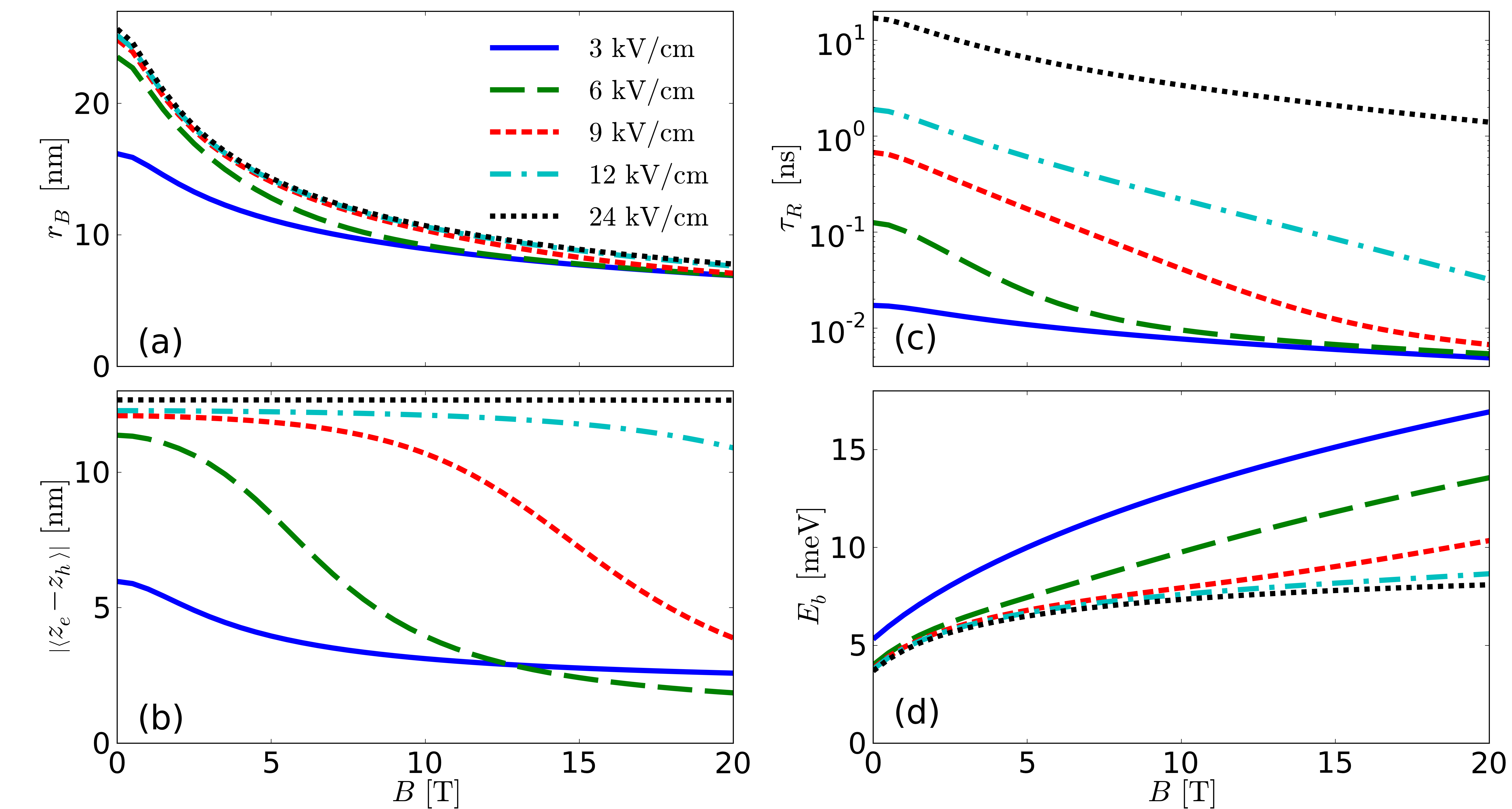}
\caption{Electric and magnetic field dependence of (a) the exciton ground state in-plane Bohr radius, (b) dipole length, (c) radiative lifetime, and (d) binding energy.}
\label{all}
\end{figure}

A number of properties of each eigenstate can be calculated from the exciton wave functions (\ref{solution}). In figure \ref{all}, we focus on the characteristics of the ground state of the system. At zero magnetic field, the results presented here agree \cite{SivalertpornError} with those of \cite{SivalertpornPRB2012}. Figure \ref{all}(a) shows the electric and magnetic field dependence of the in-plane Bohr radius, $r_B = \sqrt{\langle \rho^2 \rangle}$. Increasing the electric field increases $r_B$ due to a reduction in the {\it e-h} Coulomb interaction which is caused by their increased separation in the $z$ direction. Increasing the magnetic field has the reverse effect and shrinks the wave function in the QW plane. This shrinkage enhances the $e$-$h$ interaction which, in turn, leads to a more likely localization of the electron within the same QW as the hole and thus the ground state becomes a direct exciton. This is seen in figure \ref{all}(b) where a sharp reduction in the dipole moment $|\langle z_e - z_h \rangle |$ takes place for increasing $B$. The value of $B$ at which this transition occurs increases with $F$ as a stronger confinement of the in-plane wave function is needed to sufficiently enhance the $e$-$h$ interaction and to induce the indirect-to-direct transition. In the indirect exciton regime at high electric field ($F = 24\,{\rm kV/cm}$), the dipole moment corresponds to approximately the center-to-center distance of the QWs which is 12\,nm.

\begin{figure}
\centering
\includegraphics[width=0.7\textwidth]{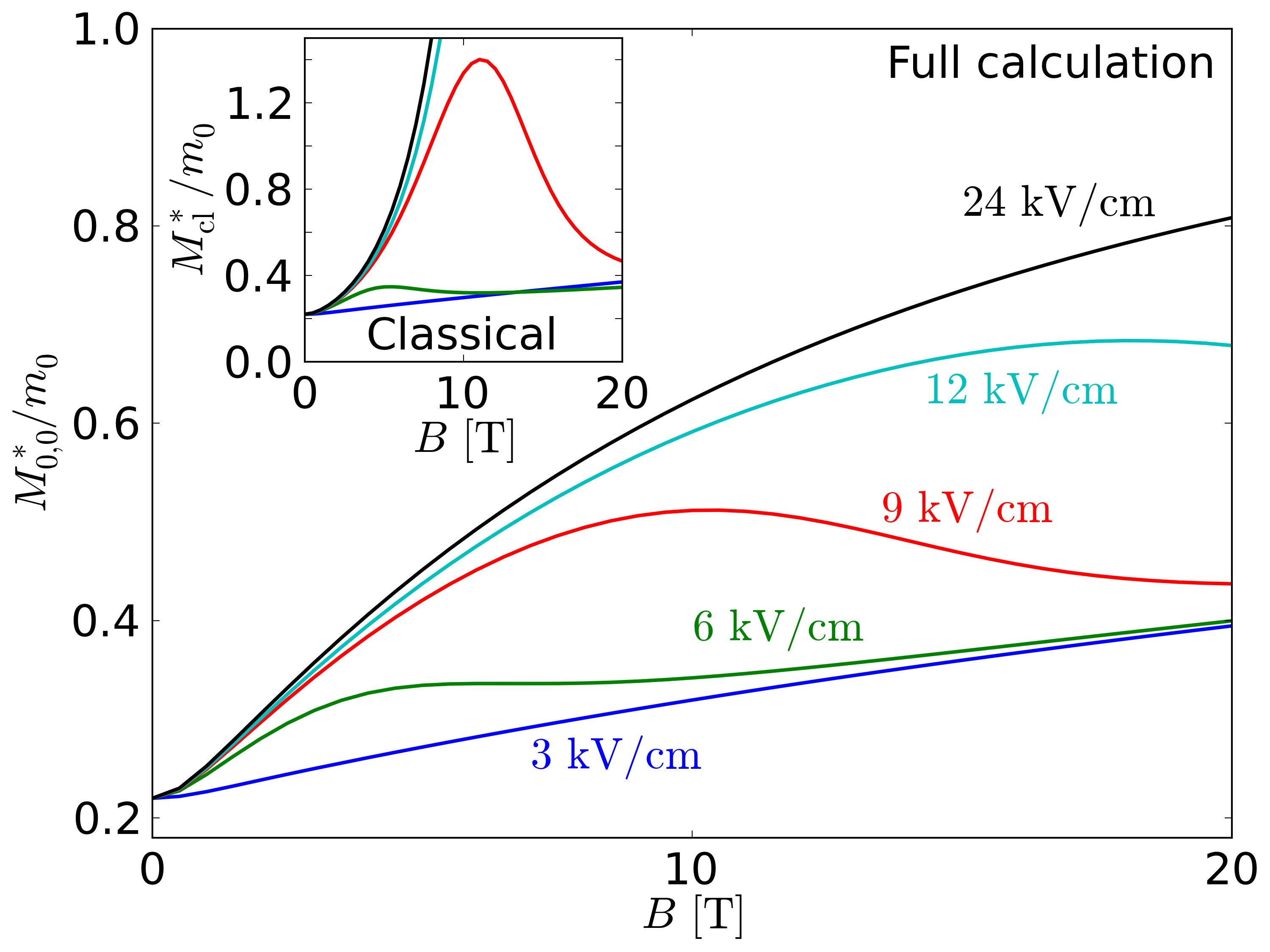}
\caption{Electric and magnetic field dependence of the exciton ground state effective mass calculated using the full MSLA. Inset: The same calculated within the classical model described in text, using the data in figures \ref{all}(a) and \ref{all}(b).}
\label{massFig}
\end{figure}

From the radiative linewidth, the radiative lifetime of the ground state exciton, $\tau_R = \hbar/(2 \Gamma_R^{(0)})$, is found. Figure \ref{all}(c) shows $\tau_R$ as a function of electric and magnetic fields. $\tau_R$ is inversely proportional to the overlap of the electron and hole wave functions (see figure \ref{zWfn}). Therefore, we find a decrease in $\tau_R$ with decreasing $F$ or increasing $B$, consistent with the shrinking of the exciton radius shown in figures \ref{all}(a) and \ref{all}(b). We note that in experiments \cite{ButovPRB1999b}, contrary to our results, the radiative lifetime of QW excitons is found to increase with increasing magnetic field. This, however, can be explained by considering the thermal distribution of excitons. Only exciton states inside the light cone, i.e. those with in-plane momentum $P < P_{\gamma}$, can decay optically. The momentum $P_{\gamma}$ corresponds to the intersection of the exciton and photon dispersions. The increase in the exciton effective mass (see figure \ref{massFig}) causes a reduction in $P_{\gamma}$ which reduces the fraction of excitons coupled to light. This increases the effective lifetime of the exciton gas---the quantity which is measured in experiments via the decay rate of the photoluminescence intensity \cite{HammackPRB2009}.

Figure \ref{all}(d) shows the binding energy $E_b = E_1^{(0)} - E_{0,0} + \hbar e B/(2 \mu c)$ where $E_{0,0}$ is the calculated exciton ground state energy in magnetic field, $E_1^{(0)} = {E}^e_{1} + {E}^h_{1}$ is the energy of the unbound {\it e-h} pair state in zero magnetic field, and the last term takes into account the Landau quantization energy of the free pair in magnetic field. The decrease in the size of the exciton, caused by either increasing $B$ or decreasing $F$, leads to an enhancement of the $e$-$h$ interaction \cite{PeetersPRB1991}. This causes a tighter binding of the exciton and, therefore, an increase in $E_b$.

The effective exciton mass, $M^\ast_{k,m}$, is calculated using second order perturbation theory (see Section \ref{massRenorm}). Figure \ref{massFig} shows the electric and magnetic field dependence of the effective mass, $M^\ast_{0,0}$, for the exciton ground state. We identify two limiting cases: Firstly, for small electric fields ($F < {\rm 3 \, kV/cm}$) where the exciton ground state is predominantly direct, there is a monotonic increase of the effective mass with increasing magnetic field. The rate of this increase is weakly dependent on the electric field. Secondly, for electric fields greater than ${\rm 3 \, kV/cm}$, there is an initial increase in the effective mass with increasing magnetic field, followed by an eventual decrease. At higher magnetic fields, the effective mass asymptotically approaches the direct exciton effective mass.

To interpret the data presented in figure \ref{massFig}, we use the classical analogy introduced in Section \ref{intro}. The exciton is approximated as two oppositely charged masses at an equilibrium separation $a_B = \sqrt{r_B^2 + \langle z_e - z_h \rangle^2}$ with center of mass motion perpendicular to the magnetic field. The Lorentz force $F_L=eBv/c$ acts on each particle to separate the charges but is balanced by the Coulomb restoring force $F_C$ that is approximately linear for small changes in separation (corresponding to small velocity and small Lorentz force): $F_L=F_C\approx k_R \Delta \rho$. In this picture, one derives from the energy dependence on the exciton velocity $v$, $M_{\rm x} v^2/2+k_R (\Delta \rho)^2/2$,  an estimate of the effective mass, $M^\ast_{\rm cl} = M_{\rm x} + e^2B^2/(c^2k_R)$. The `spring constant' $k_R$ is the coefficient of the restoring force. It can be approximated via linearization of the slope in the Coulomb potential at $\rho = r_B$ which gives
\begin{equation}
k_R = -\left[\frac{\partial^2}{\partial \rho^2} V_c (\sqrt{\rho^2 + \langle z_e - z_h \rangle^2})\right]_{\rho = r_B}.
\end{equation}
The inset in figure \ref{massFig} shows this classical calculation using the parameters $r_B$ and $\langle z_e - z_h \rangle^2$ characterizing the exciton taken from the full calculation, see figures \ref{all}(a) and \ref{all}(b). The curves for $F=3$\,kV/cm and 6\,kV fully describe the main trends seen in the full calculation and even demonstrate to some extent a  quantitative agreement. For larger electric and magnetic fields this classical model fails because the spring constant $k_R$ becomes too small and the linearization of the restoring force used in the model no longer works.

The classical model allows us to understand the exciton mass dependence in figure \ref{massFig}. Indeed, from the calculation presented in figure \ref{all}, we find a greater $k_R$ for direct excitons than indirect excitons. Qualitatively, this is because the indirect exciton has a weaker binding due to the increased spatial separation of the electron and hole. The difference in $k_R$ leads to a greater rate of effective mass increase with magnetic field for indirect excitons than for direct excitons. This explains the effective mass dependence in the low and high electric field limits in figure \ref{massFig}. At intermediate fields ($6-12\,{\rm kV/cm}$), the initial increase and then decrease of the effective mass is a consequence of the ground state tending from an indirect to a direct exciton. This happens due to a shrinkage of the exciton wave function caused by the magnetic field. In turn, the {\it e-h} Coulomb attraction is enhanced, making the direct state lower in energy than the indirect state.

\section{Comparison with the 2D limit}
\label{2Dcomparison}
In order to verify our model, we make comparisons with previous results in which the limiting case of 2D excitons were treated. To make this comparison, we include just one electron and one hole state [$N_e=N_h = 1$ in (\ref{solution})] taking the wave functions in the form $|{\psi}^{e}(z)|^2 = \delta(z)$ and $|{\psi}^{h}(z)|^2 = \delta(z-d_z)$,  where $\delta(z)$ is the Dirac delta function and $d_z$ is the nominal distance between the QWs. Technically, this is done using normalized Gaussian functions for ${\psi}^{e,h}(z_{e,h})$, for the electron and hole centered on $0$ and $d_z$, respectively, and taking the limit of their widths to zero.

\begin{figure}
\centering
\includegraphics[width=0.7\textwidth]{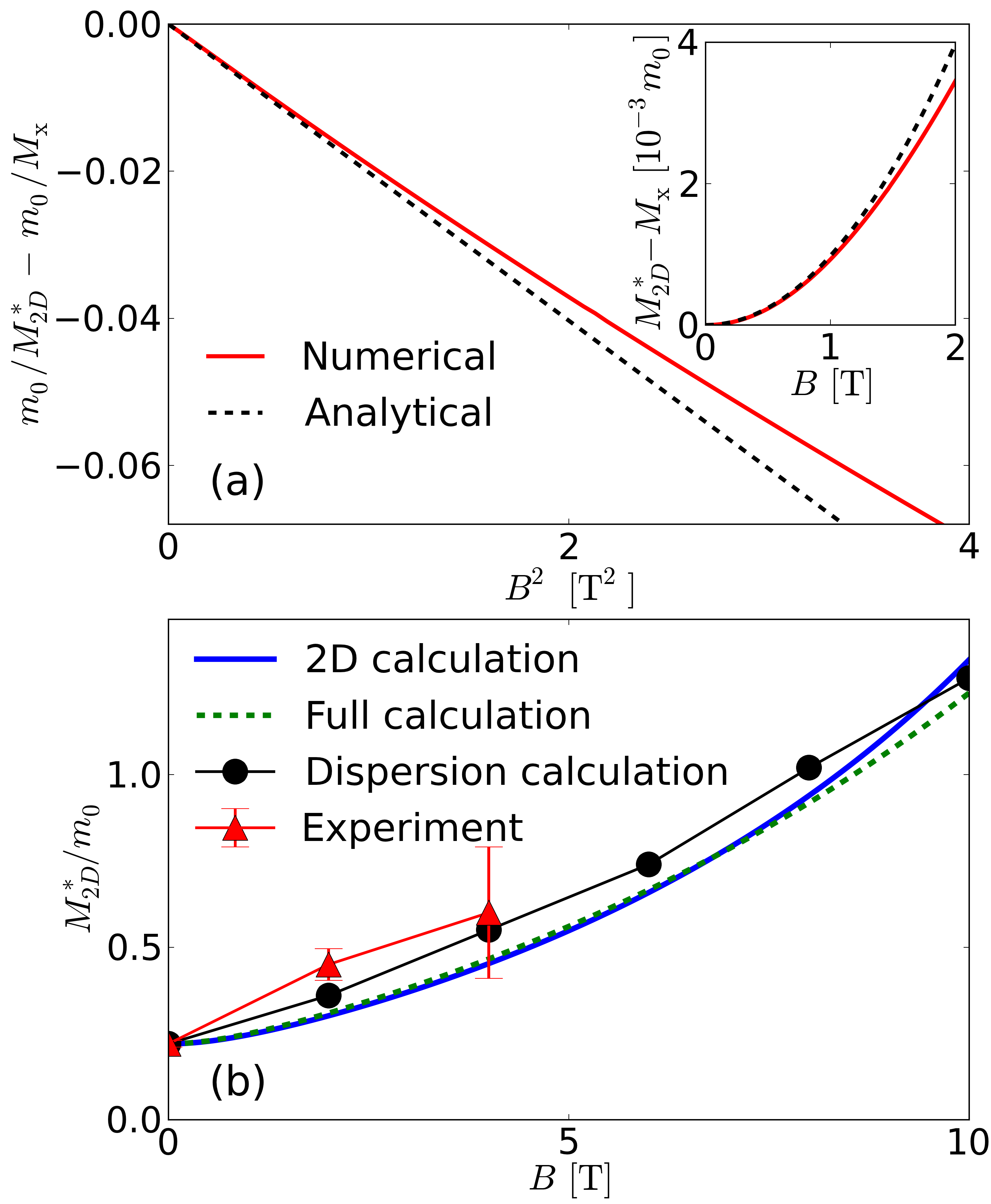}
\caption{(a) Present numerical calculation (solid red line) and analytic calculation of \cite{ArseevJETP1998} using (\ref{DzyubenkoMass}) (dashed black line) of the inverse of the mass renormalization against $B^2$ for 2D excitons in a single QW. The inset shows the corresponding effective mass increase as a function of $B$. (b) Indirect exciton effective mass as a function of $B$, as determined by the full calculation (green dashed line), using 2D electron and hole wave functions separated by $d_z = 11.5\,{\rm nm}$ (blue line), via the dispersion relation calculated in \cite{LozovikPRB2002} (black circles with a line) and experimentally \cite{ButovPRL2001} (red triangles with error bars and a line). The error bars show the errors extracted from the data presented in \cite{ButovPRL2001}.}
\label{2Dmass}
\end{figure}

First, we consider the mass renormalization of 2D direct excitons in single QWs ($d_z=0$). This was calculated by Arseev and Dzyubenko \cite{ArseevJETP1998}. The analytical expression for the exciton mass, which was derived by treating the magnetic field as a perturbation, was given as \cite{DzyubenkoError}
\begin{equation}
\frac{1}{M^\ast_{\rm 2D}} = \frac{1}{M_{\rm x}}\left[1 - \frac{42 \mu}{16^2 M_{\rm x}}\left( \frac{a_{\rm x}}{l_B}\right)^4\right].
\label{DzyubenkoMass}
\end{equation}
This is relevant in the low magnetic field limit where the magnetic length is much greater than the exciton Bohr radius, $l_B = \sqrt{\hbar c / e B} \gg a_{\rm x} = \varepsilon \hbar^2/ \mu e^2$. For the considered system, this corresponds to $B \ll 2.6 \, {\rm T}$. Figure \ref{2Dmass}(a) shows the inverse of the effective mass renormalization as a function of $B^2$ calculated numerically via (\ref{effectiveMass}) and using the analytic expression (\ref{DzyubenkoMass}). There is a strong agreement between the different approaches for small $B$. For increasing $B$ above 1T, the results depart as expected, since the magnetic length approaches the Bohr radius.

We also compare our results with the work of Lozovik \cite{LozovikPRB2002} which treated the case of 2D indirect excitons in CQWs. In that approach, the exciton Schr\"odinger equation was solved with a finite center of mass momentum that enabled calculation of the dispersion relation. The effective mass was then extracted from a parabolic fit to the bottom of the dispersion band. The mass calculated by this method is shown in figure \ref{2Dmass}(b). To make a suitable comparison, we calculate the 2D effective mass using the same parameters as in \cite{LozovikPRB2002} and \cite{ButovPRL2001}: $d_z = 11.5\,{\rm nm}$, $\mu = 0.049\,m_0$, $\varkappa = 0.11\,m_0$, and $\varepsilon = 12.1$. The full calculation using the same parameters with $F = 24\,{\rm kV/cm}$ is shown for reference.

In figure \ref{2Dmass}(b), we also show the effective mass dependence on magnetic field as was determined experimentally \cite{LozovikPRB2002,ButovPRL2001} using an in-plane component in the magnetic field to map the indirect exciton dispersion by shifting it with respect to the photon cone. We note that the quantitative agreement between theory and experiment could be improved by further adjustment of the parameters $\mu$ and $\varkappa$. The calculated mass is quite sensitive to these parameters which are not known to great precision. Other possible sources of discrepancy could be the inhomogeneous permittivity, non-parabolicities of the bands and valence band mixing. One should also consider the limited accuracy of the effective mass measurement which entails fitting a parabola to an almost flat dispersion. The dispersion curve is extracted from the energy of the exciton line in the emission spectrum. This can be influenced by the disorder intrinsic to the CQW sample and the density dependent blue shift due to the exciton-exciton interaction. On that basis, our data is in good agreement with the experimentally measured exciton effective mass and the combination of the effects that we neglect is rather minor.

\section{Conclusions}
\label{conclusions}
We have studied the combined effect of perpendicular electric and magnetic fields on exciton states in coupled quantum wells. This was done using the rigorous multi-sub-level approach (MSLA) to solve the full exciton Schr\"odinger equation in the effective mass approximation. In this approach, the exciton wave function is expanded into Coulomb uncorrelated electron-hole pair states and found using a matrix generalization of the shooting method with high stability and accuracy provided by Numerov's algorithm. We calculated the optical transition energy and linewidths for different exciton states and obtained the electric and magnetic field dependence of the exciton absorption spectrum. For the exciton ground state, we evaluated the in-plane Bohr radius, static dipole moment, radiative lifetime and binding energy. Furthermore, we have calculated the magnetic field-induced exciton mass renormalization in arbitrary electric and magnetic field, by treating the exciton momentum as a small perturbation. In particular, our calculations demonstrate a non-monotonous behavior of the exciton effective mass in magnetic field which is understood in terms of the direct-to-indirect exciton transition and also illustrated in a simple classical model. Our results for the quantum well width tending to zero are in good agreement with different 2D approaches available in the literature. However, in contrast to the 2D approaches, the full MSLA captures the direct-to-indirect transitions of the exciton ground state for varying electric and magnetic fields which are necessary for a full and detailed description of the system.

\ack
We would like to thank A B Dzyubenko for helpful discussions and L V Butov for reading the manuscript. Support of this work by the EPSRC (grant EP/L022990/1) is gratefully acknowledged. This work was performed using the computational facilities of the ARCCA Division, Cardiff University.

\appendix
\section{Exciton Hamiltonian in a CQW in perpendicular electric and magnetic fields}
\label{appA}

Using the factorizable form of the wave function (\ref{COM1}-\ref{COM2}) we make a transformation of the full Hamiltonian (\ref{Ham})
\begin{equation}
\hat{H}^{\bf P}_{\rm x}({\bf r}_e,{\bf r}_h) = \chi_{\bf P}^{-1} \hat{H}(\boldsymbol\rho,z_e,z_h) \chi_{\bf P}\,,
\end{equation}
removing the in-plane center of mass coordinate from the problem. The exciton Hamiltonian then takes the form
\begin{eqnarray}
\hat{H}^{\bf P}_{\rm x}(\boldsymbol\rho,z_e,z_h) &=& \hat{H}^\perp_e(z_e) + \hat{H}^\perp_h(z_h) + \hat{W}^{\bf P}_B(\boldsymbol\rho) + E_g \nonumber \\
 &&+ V_C\left(\sqrt{\rho^2 + (z_e - z_h)^2}\right)\,,
\label{Exc-Ham}
\end{eqnarray}
where $\hat{H}^\perp_{e,h}$ and $V_C$ are given by (\ref{Heh}) and 
\begin{equation}
\hat{W}^{\bf P}_{B}(\boldsymbol\rho) = \chi_{\bf P}^{-1} \hat{H}_B (\boldsymbol\rho_e,\boldsymbol\rho_h) \chi_{\bf P}
\end{equation}
with
\begin{eqnarray}
\hat{H}_B(\boldsymbol\rho_e,\boldsymbol\rho_h) &=& \frac{\hbar^2}{2m_e^\parallel}\left(-\nabla_e^2 - 2 i\gamma  {\bf A}_e \!\cdot\! \nabla_e + \gamma^2 A^2_e \right) \ \ \ \ \ \ \ \ \nonumber \\
&\,\ +\!\!\!& \frac{\hbar^2}{2m^\parallel_h}\left(-\nabla_h^2 + 2 i\gamma {\bf A}_h \!\cdot \!\nabla_h + \gamma^2 A^2_h \right).
\label{HB}
\end{eqnarray}
For brevity of notation, we have introduced in (\ref{HB})  $\gamma = e/\hbar c$, ${\bf A}_{e,h} = {\bf A}(\boldsymbol\rho_{e,h})$, and $\nabla_{e,h} = \nabla_{\boldsymbol\rho_{e,h}}$, where $\nabla_{\boldsymbol\rho}$ is a 2D nabla-operator. For the same reason, we also drop below the index ${\bf P}$ in $\chi_{\bf P}$, ${\rm x}$ in $M_{\rm x}$, and $\parallel$ in $m_{e,h}^\parallel$. Taking an arbitrary function $f(\boldsymbol\rho)$ we obtain from (\ref{HB}):
\begin{eqnarray}
\hat{H}_B \chi f &=& \frac{\hbar^2}{2m_e}\Bigl[ -\chi \nabla_e^2 f - 2 \nabla_e \chi \cdot \nabla_e f - f \nabla_e^2 \chi  \ \ \ \ \ \ \ \ \ \nonumber\\
&& - 2i \gamma {\bf A}_e \!\cdot\! (\chi \nabla_e f + f\nabla_e \chi) + \gamma^2 A_e^2 \chi f \Bigr] \nonumber \\
&&+ \frac{\hbar^2}{2m_h}\Bigl[ -\chi \nabla_h^2 f - 2 \nabla_h \chi \cdot \nabla_h f - f \nabla_h^2 \chi  \nonumber \\
&& + 2i \gamma {\bf A}_h\! \cdot\! (\chi \nabla_h f + f \nabla_h \chi) + \gamma^2 A_h^2 \chi f \Bigr]\,.
\label{A6}
\end{eqnarray}
Noting that $\nabla_e f = \nabla f$ and $\nabla_h f = -\nabla f$, (\ref{A6}) can be written as
\begin{eqnarray}
\hat{H}_B \chi f = \frac{\hbar^2}{2} \left[-\frac{\chi}{\mu}\nabla^2 f - 2 \nabla f\cdot \left(\frac{\nabla_e \chi}{m_e} - \frac{\nabla_h \chi}{m_h}\right)\right. \ \ \ \ \ \ \ \ \\
-f \left(\frac{\nabla_e^2 \chi}{m_e} + \frac{\nabla_h^2 \chi}{m_h}\right) - 2i \gamma \frac{{\bf A}_e}{m_e} \cdot (\chi \nabla f+ f \nabla_e \chi) \\ \left. + 2 i \gamma \frac{{\bf A}_h}{m_h} \cdot (-\chi \nabla f + f \nabla_h \chi)+ \gamma^2 \left(\frac{A_e^2}{m_e}+\frac{A_h^2}{m_h}\right) \chi f \right].
\end{eqnarray}
Finally, noting that 
\begin{eqnarray}
\nabla_{e,h} {\bf A}(\boldsymbol\rho) \cdot  {\bf R}\! &=&\! \frac{1}{2}\nabla_{e,h}{\bf B}\times (\boldsymbol\rho_e\!-\!\boldsymbol\rho_h)\!\cdot\!\left(\frac{m_e}{M}\boldsymbol\rho_e\!+\!\frac{m_h}{M}\boldsymbol\rho_h\right) \nonumber\\
&=&\!\frac{1}{2}\nabla_{e,h}{\bf B}\cdot \boldsymbol\rho_e\times\boldsymbol\rho_h =-{\bf A}_{h,e}\,,
\end{eqnarray}
obtain
\begin{equation}
\nabla_{e,h} \chi = i \left( \frac{m_{e,h}}{M\hbar}{\bf P} \mp \gamma {\bf A}_{h,e} \right) \chi, 
\end{equation}
and after some algebraic simplifications using ${\bf A}(\boldsymbol\rho) = {\bf A}_e - {\bf A}_h$,  find the required term $\hat{W}^{{\bf P}}_B$ of the exciton Hamiltonian (\ref{Exc-Ham}):
\begin{eqnarray}
\hat{W}^{{\bf P}}_B (\boldsymbol\rho) &=& -\frac{\hbar^2}{2 \mu} \nabla^2 - \frac{i e \hbar}{\varkappa c}{\bf A}(\boldsymbol\rho) \cdot \nabla + \frac{e^2}{2\mu c^2}A^2(\boldsymbol\rho) \nonumber \\
&&+ \frac{P^2}{2M}+ \frac{2e}{Mc} {\bf P} \cdot {\bf A}(\boldsymbol\rho)\,.
\label{wHam_B}
\end{eqnarray}
This term is the only one in (\ref{Exc-Ham}) which carries in the dependence on the magnetic field $B$ and the exciton in-plane momentum ${\bf P}$. For the exciton wave function in the form of (\ref{COM1}), with the split-off angular motion, the exciton Hamiltonian is further reduced to (\ref{mainHamiltonian}) for ${\bf P}=0$  and to (\ref{pert}) for ${\bf P}\neq0$.

\section{Perturbative calculation of the exciton mass enhancement due to a magnetic field}
\label{appB}

We treat the exciton center of mass momentum ${\bf P}$ as a perturbation and first seek solutions in a finite magnetic field with ${\bf P} = 0$. 
Without loss of generality, we can take ${\bf P}$ along the $y$-axis, which gives ${\bf P} \cdot {\bf A}(\boldsymbol\rho) = \frac{1}{2} B P \rho \cos (\phi)$ in polar coordinates, and evaluate the matrix elements,
\begin{eqnarray}
I^{m,s}_{k,j} = \frac{1}{P} \left\langle k, m \left| \frac{2e}{M_{\rm x}c} {\bf P} \cdot {\bf A}(\boldsymbol\rho) \right| m+s, j \right\rangle \nonumber \\
= \frac{eB}{M_{\rm x}c} \int \int_{-\infty}^{\infty}\!\! dz_e \, dz_h  \int_0^{2 \pi}\!\!d \phi \int_0^\infty \!\!\rho \, d \rho\,\rho \cos(\phi)e^{i s \phi} \nonumber \\
\times \sum_n \Phi_n(z_e,z_h) {\phi}_{n}^{k,m}(\rho) \sum_{n'} \Phi_{n'}(z_e,z_h) {\phi}_{n'}^{j,m+s}(\rho)\,.
\end{eqnarray}
We make use of the following identities:
\begin{eqnarray}
\int_0^{2 \pi} e^{i m \phi} \cos(\phi) \, d\phi = \pi \delta_{m^2,1} \\
\int_{-\infty}^{\infty} {\psi}^e_{p_n}(z_e) {\psi}^e_{p_{n'}}(z_e) \, dz_e = \delta_{p_n,p_{n'}}, \\
\int_{-\infty}^{\infty} {\psi}^h_{q_n}(z_h) {\psi}^h_{q_{n'}}(z_h) \, dz_h = \delta_{q_n,q_{n'}}, \\
\delta_{p_n,p_{n'}}\delta_{q_n,q_{n'}} = \delta_{n,n'}.
\end{eqnarray}
The matrix elements then simplify to
\begin{equation}
I^{m,s}_{k,j} = \delta_{s^2,1}\,\frac{\pi e B}{M_{\rm x} c}\sum_n \int_0^\infty {\phi}_{n}^{k,m}(\rho) {\phi}_{n}^{j,m+s}(\rho) \rho^2 \, d\rho\,.
\end{equation}
The correction, $M_{k,m}^{(B)}$, to the effective mass of exciton state $|k,m\rangle$ is then given by the 2nd perturbation order:
\begin{equation}
\frac{1}{2 M_{k,m}^{(B)}} = \sum_{j} \left[\frac{(I_{k,j}^{m,-1})^2}{E_{k,m} - E_{j,m-1}} + \frac{(I_{k,j}^{m,1})^2}{E_{k,m} - E_{j,m+1}}\right]\,.
\label{series}
\end{equation}
Finally, the renormalized effective mass $M^\ast_{k,m}$ of an exciton in a magnetic field is determined by
\begin{equation}
\frac{1}{M_{k,m}^\ast} = \frac{1}{M_{\rm x}} + \frac{1}{M_{k,m}^{(B)}}.
\end{equation}

\begin{figure}
\centering
\includegraphics[width=0.7\textwidth]{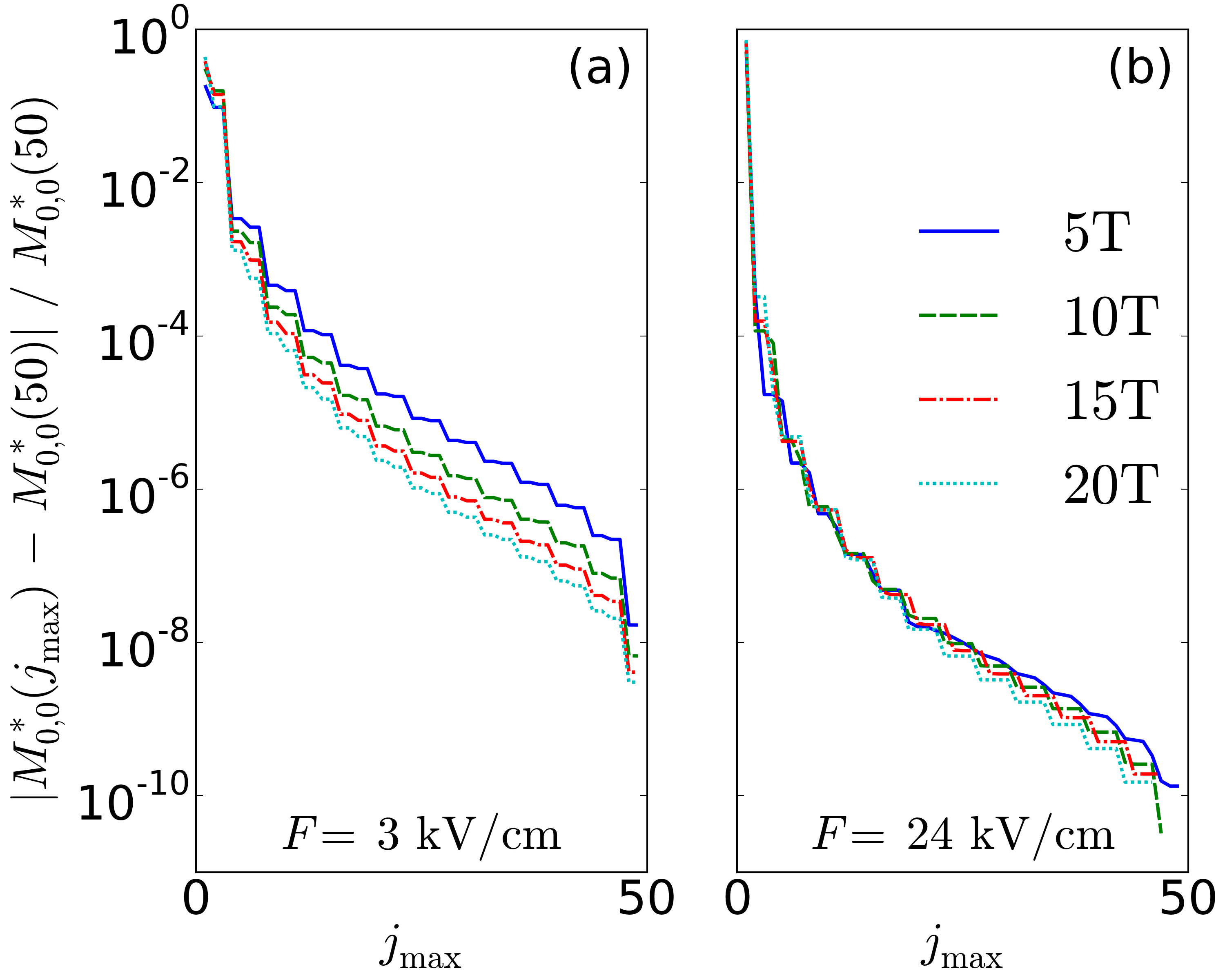}
\caption{Relative error in the exciton ground state mass calculation via (\ref{series}) as a function of $j_{\rm max}$, for different values of electric and magnetic fields.}
\label{convergence}
\end{figure}

Figure \ref{convergence} shows some illustrative examples of the convergence of the effective mass with respect to $j_{\rm max}$ for $k=m=0$. Here, $j_{\rm max}$ is the number of states taken into account in the perturbation series, (\ref{series}). It can be seen that the series converges rapidly for the range of electric and magnetic fields considered here. $j_{\rm max} = 15$ was used in all data except figure \ref{convergence}, achieving a relative error in the effective mass of less than $10^{-4}$.

\end{document}